\documentclass[%
reprint,
superscriptaddress,
amsmath,amssymb,longbibliography,
aps,prb
]{revtex4-2}
\usepackage[utf8]{inputenc}
\usepackage[T1]{fontenc} 
\usepackage{graphicx}
\usepackage{dcolumn}
\usepackage{bm}
\usepackage{amsmath}
\usepackage{lmodern}
\usepackage{mathtools}
\usepackage{float}
\usepackage[x11names]{xcolor}
\definecolor{PRX}{RGB}{46,48,146}
\definecolor{MyBlack}{RGB}{35,31,32}
\usepackage[colorlinks,citecolor=PRX  ,linkcolor=PRX,urlcolor=PRX]{hyperref}
\usepackage{siunitx}
\usepackage{chemformula}
\usepackage{upgreek}

\usepackage{chemformula} 
\usepackage[T1]{fontenc} 
\usepackage{soul}

\usepackage[normalem]{ulem}

\usepackage[normalem]{ulem}

\usepackage{titlesec}
\titleformat{\section}[display]{\bf\large}{}{0pt}{}
\titleformat{\subsection}[display]{\bf}{}{0pt}{}
\titlespacing\section{0pt}{12pt plus 4pt minus 2pt}{0pt plus 2pt minus 2pt}
\titlespacing\subsection{0pt}{12pt plus 4pt minus 2pt}{0pt plus 2pt minus 2pt}
\titlespacing\subsubsection{0pt}{12pt plus 4pt minus 2pt}{0pt plus 2pt minus 2pt}

\begin{document}

\title{Robust Room-Temperature Polariton Condensation and Lasing in Scalable FAPbBr\textsubscript{3} Perovskite Microcavities}

\author{Mateusz Kr\'{o}l}
\thanks{These authors contributed equally}
\affiliation{ARC Centre of Excellence in Future Low-Energy Electronics Technologies and Department of Quantum Science and Technology, Research School of Physics, The Australian National University, Canberra, ACT 2601, Australia}
\email{mateusz.krol@anu.edu.au}

\author{Mitko Oldfield}
\thanks{These authors contributed equally}
\affiliation{ARC Centre of Excellence in Future Low-Energy Electronics Technologies and School of Physics and Astronomy, Monash University, Clayton, VIC 3800, Australia}

\author{Matthias Wurdack}
\thanks{These authors contributed equally}
\affiliation{ARC Centre of Excellence in Future Low-Energy Electronics Technologies and Department of Quantum Science and Technology, Research School of Physics, The Australian National University, Canberra, ACT 2601, Australia}
\affiliation{Hansen Experimental Physics Laboratory, Stanford University, Stanford CA 94305}

\author{Eliezer Estrecho}
\affiliation{ARC Centre of Excellence in Future Low-Energy Electronics Technologies and Department of Quantum Science and Technology, Research School of Physics, The Australian National University, Canberra, ACT 2601, Australia}
\author{Gary\,Beane}
\affiliation{ARC Centre of Excellence in Future Low-Energy Electronics Technologies and School of Physics and Astronomy, Monash University, Clayton, VIC 3800, Australia}
\author{Yihui Hou}
\affiliation{School of Engineering, The Australian National University, Canberra, ACT 2601, Australia}
\author{Andrew G. Truscott}
\affiliation{Department of Quantum Science and Technology, Research School of Physics, The Australian National University, Canberra, ACT 2601, Australia}
\author{Agustin Schiffrin}
\affiliation{ARC Centre of Excellence in Future Low-Energy Electronics Technologies and School of Physics and Astronomy, Monash University, Clayton, VIC 3800, Australia}
\author{Elena A. Ostrovskaya}
\affiliation{ARC Centre of Excellence in Future Low-Energy Electronics Technologies and Department of Quantum Science and Technology, Research School of Physics, The Australian National University, Canberra, ACT 2601, Australia}

\begin{abstract}
Exciton-polariton condensation in direct bandgap semiconductors strongly coupled to light enables a broad range of fundamental studies and applications like low-threshold and electrically driven lasing. Yet, materials hosting exciton-polariton condensation in ambient conditions are rare, with fabrication protocols that are often inefficient and non-scalable. Here, room-temperature exciton-polariton condensation and lasing is observed in a  microcavity with embedded formamidiniumlead bromide (FAPbBr\textsubscript{3}) perovskite film. This optically active material is spin-coated onto the microcavity mirror, which makes the whole device scalable up to large lateral sizes. The sub-$\upmu$m granulation of the polycrystalline FAPbBr\textsubscript{3} film allows for observation of polariton lasing in a single quantum-confined mode of a polaritonic `quantum dot'. Compared to random photon lasing, observed in bare FAPbBr\textsubscript{3} films, polariton lasing exhibits a lower threshold, narrower linewidth, and an order of magnitude longer coherence time. Both polariton and random photon lasing are observed under the conditions of pulsed optical pumping, and persist without significant degradation for up to $6$ and $17$ hours of a continuous experimental run, respectively. This study demonstrates the excellent potential of the FAPbBr\textsubscript{3} perovskite as a new material for room-temperature polaritonics, with the added value of efficient and scalable fabrication offered by the solution-based spin-coating process.
\end{abstract}

\maketitle

\begin{figure*}
  \includegraphics{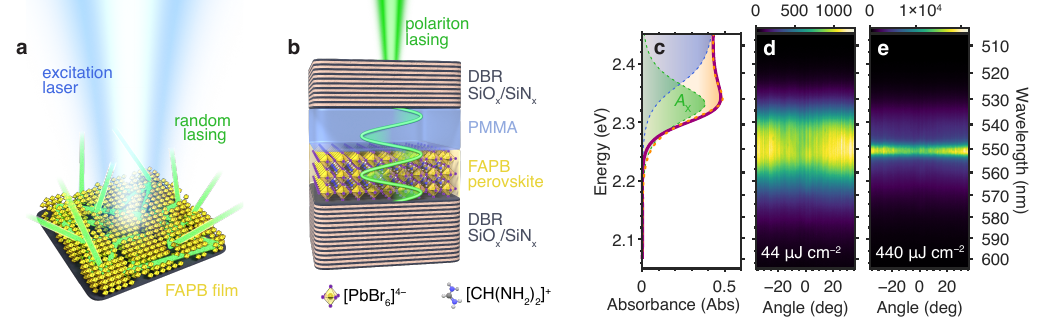}
\caption{\textbf{a,} Schematic of random lasing in bare FAPB perovskite film. \textbf{b,} Schematic of the microcavity structure. \textbf{c,}\,Absorbance of bare FAPB film. Experimental results (purple solid curve) were fitted (dashed orange curve) using Elliott's theory with excitonic (green dashed curve) and free carrier contributions (blue dashed curve). \textbf{d, e,} Angle resolved photoluminescence (PL) spectra for excitation fluences below (d) and above (e) the lasing threshold.}
  \label{fig:Fig1}
\end{figure*}

\section{Introduction}
Strong light-matter coupling in direct bandgap semiconductors gives rise to exciton polaritons -- hybrid bosonic quasiparticles composed of excitons (electron-hole pairs) dressed by photons \cite{Weisbuch1992,Houdre1994}. Exciton polaritons have a small effective mass inherited from photons and can undergo bosonic condensation in much less demanding temperature and pressure regimes compared to other massive bosons \cite{Deng2002,Kasprzak2006,Balili2007,Christopoulos2007,Wertz2010}. Stability of excitons at elevated temperatures, and the possibility to achieve the density-driven transition to condensation with relatively low optical pumping powers has led to the explosion of research into polariton-hosting materials \cite{Schneider2018,Ardizzone2019,Su2021,Zhao2021,anton2021bosonic}, fundamental properties of polariton condensates \cite{Carusotto2013}, and prototype polaritonic devices \cite{Sanvitto2016}. Amongst the latter, the concept of a polariton laser \cite{Deng2003} that does not require population inversion and derives its light emission from the decay of the macroscopic quantum coherent condensate into coherent photons, has attracted considerable attention due to extremely low thresholds \cite{Su2017,Zhao2021} and the possibility of electrical injection \cite{Hofling2013,Das2014}.
\medskip

A prerequisite for real-world applications is that polariton condensation and lasing occur at room-tem\-pe\-ra\-tu\-re -- a requirement achieved only in some materials, e.g., GaN \cite{Christopoulos2007}, ZnO \cite{Lu2012}, organics \cite{Plumhof2013} and lead halide perovskite films \cite{Su2021,Bao2022,BaoNMat2022}. Perovskites, in particular, are attracting growing attention due to their relative ease of fabrication and high rate of exciton production (quantum efficiency) when the material is exposed to light \cite{Su2021}. However, further progress towards room-temperature polaritonic devices critically depends on the stability of the optically active materials against prolonged exposure to the ambient environment and light \cite{Liu:23}, as well as on the development of a reliable and scalable process of encapsulating them inside an optical microcavity. The latter is essential to create a confined photon mode with a high quality factor, which not only enables strong light-matter coupling, but also determines the lifetime of the exciton polariton, as well as affects the coherence properties of the condensate and the resulting laser. The process of encapsulating the exciton-hosting material in a microcavity typically involves sandwiching the material between two high-reflectivity distributed Bragg reflector (DBR) mirrors. The limited lateral size of the active material itself, and/or the need to avoid fabrication steps that could degrade the intrinsic optical properties (i.e. emission and absorption) of the material, often results in a small functional area of the assembled microcavity. This limits the scalability of polaritonic devices and motivates the search for novel exciton-hosting materials that can be synthesized on a wafer scale, do not degrade when exposed to light and chemicals, and can be readily integrated into multi-layer heterostructures for room-temperature polaritonics.
\medskip

Herein, we report on the observation of room-temperature polariton condensation in a formamidinium lead bromide (FAPbBr\textsubscript{3}, FAPB) perovskite film and characterize the lasing properties of this material in both the photonic and polaritonic regimes. Owing to the excellent thermal, photo- and chemical stability of FAPB \cite{Borchert2017,Leijtens2017}, as well as its remarkable optoelectronic properties, such as high quantum efficiency, this material has emerged as a strong candidate for light-emitting applications. Importantly for polaritonics, FAPB excitons have a binding energy ($E_B=31.6\pm0.9$\,meV) above $k_\mathrm{B} T$ at room temperature, and a large oscillator strength, which enables strong light-matter coupling, formation of exciton-polaritons and polariton condensation in ambient conditions.

\section{Results and Discussion}

\subsection{Material Synthesis and Sample Fabrication}

\begin{figure*}
  \includegraphics{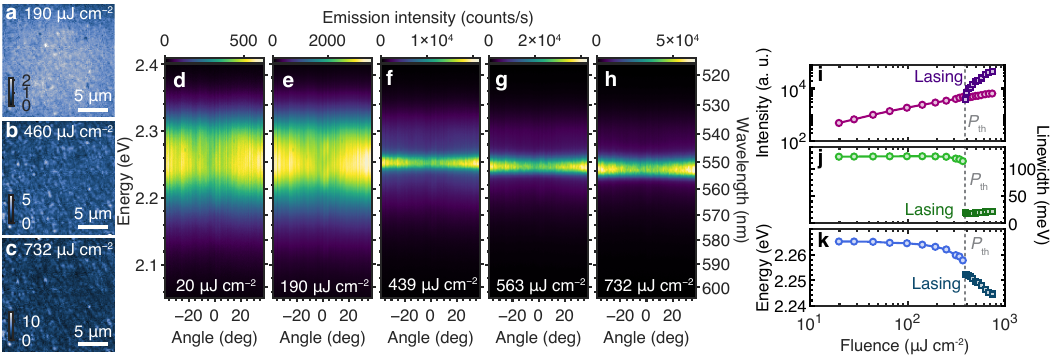}
\caption{\textbf{a--c,} Map of real-space emission from bare FAPB perovskite film, for different excitation fluences below (a) and above (b, c) the random lasing threshold. \textbf{d--h,} Angle-resolved PL spectra  for increasing excitation fluence. \textbf{i--k,} Maximum emission intensity, spectral linewidth and spectral peak position as a function of the excitation fluence. Circles correspond to spontaneous emission while squares to the random lasing signal. Vertical dashed line marks the lasing threshold.}
  \label{fig:Fig2}
\end{figure*}

The FAPB perovskite is an ideal candidate for exciton-polariton formation and condensation in spin-coated microcavities. This is due to the optimal absorption of FA-based perovskites at the excitonic resonance \cite{Mannino2020,Shivarudraiah2021}, which enables strong exciton-photon interactions. Moreover, high mobility, long charge carrier lifetimes and substantial diffusion lengths \cite{Zhumekenov2016} in these materials promote strong exciton-exciton interactions. In addition, FAPB excitons have a binding energy on the order of, or above, $k_\mathrm{B}T$ at room temperature \cite{Galkowski2016,Geng2023,Bokdam2016}. The chemical, thermal and photo-stability of FAPB \cite{Borchert2017}, as well as the low defect density \cite{Oranskaia2018}, makes it extremely promising for integration into optical microcavities and studies of strong light-matter coupling under optical pumping conditions. To achieve strong coupling, FAPB must be embedded within a high quality factor (Q-factor) microcavity. This requires an active medium with a constant thickness, minimal surface roughness, and high crystallinity. Spacer layers within the microcavity must also satisfy the former two conditions. 

Section S1 in Supporting Information (SI) contains details of the synthesis parameters that were used to produce high-quality FAPB thin films of optimal thickness (dictated by the $5\lambda/2$ active layer requirement of the microcavity) and minimal roughness, the fabrication of the cavity spacer, and the spin-coating technique that was used to encapsulate both the active layer and the spacer between the cavity mirrors.

The key feature of the material synthesis is spin-coating of the precursor solution deposited on the bottom DBR stack on a fused silica (FS) substrate. Subsequent antisolvent-aided crystallization and thermal annealing (to promote sintering) leads to heterogeneous nucleation and formation of high-quality polycrystalline films with uniform coverage of $\sim 500$ nm thickness and root-mean-squared (RMS) surface roughness of $16$ nm, as determined by atomic force microscopy (AFM) and  contact profilometry (Section S2 in SI). The spin-coated polycrystalline films have a granular nature that can be resolved by AFM (Fig.\,S1 in SI), with individual grains of $500$--$1000$\,nm size. The grain boundaries affect the exciton diffusion and non-radiative recombination rates, the latter being low for larger grain sizes \cite{Nie:2015}.
\medskip

A polymethyl methacrylate (PMMA) layer spin-coated on top of the FAPB film (see Section S1 in SI) acted as a cavity spacer and protection \cite{Yun2022_microcavity} of the active material against damage during the top DBR deposition. This PMMA layer also served to further reduce the RMS surface roughness to $8-9$ nm (at $0.16$ $\upmu$m lateral resolution, see SI, Fig. S2), resulting in the overall improvement of the cavity Q-factor.
\medskip

The encapsulation of the active material in an optical microcavity for strong light-matter coupling was subsequently achieved by depositing a top DBR mirror onto the PMMA-capped FAPB. Both the top and the bottom DBR stacks were fabricated with alternating $\lambda/4$ layers of silicon nitride (SiNx) and silicon oxide (SiOx) deposited by plasma-enhanced chemical vapour deposition (see SI). 
\medskip

Two sets of samples were fabricated using the above process in order to compare the photon lasing produced by FAPB excitons and polariton lasing produced by condensed exciton polaritons. The first set contained PMMA encapsulated and thermally annealed FAPB films spin-coated on the bottom DBR with a quartz substrate, referred to as `bare' FAPB in Fig. \ref{fig:Fig1}a. The second set of samples contained the FAPB films embedded in a full DBR microcavity with a PMMA spacer, as shown in Fig.\,\ref{fig:Fig1}b. Out of the total 22 microcavity samples that demonstrated strong exciton-photon coupling and formation of exciton polaritons, 7 were chosen for further investigation of polariton condensation and lasing. The experimental data below is representative of the typical behavior of bare FAPB and microcavity samples, and correspond to measurements performed on the same sample of each type. All measurements were performed under the conditions of optical excitation by a pulsed fs-laser (parameters detailed in SI, Section S3).

\subsection{Exciton absorption and emission}

The absorbance spectrum for the bare FAPB is shown in Fig. \ref{fig:Fig1}c. It exhibits an absorption onset at ${\sim}2.3$\,eV, above which a monotonic increase and saturation in absorption is observed as a function of photon energy. This behavior can be accurately captured by Elliot’s theory \cite{Elliott1957}, where the total energy-dependent absorbance, $A=A_{\text{CC}}+A_\text{X}$, includes both charge carrier ($A_{\text{CC}}$) and exciton contributions ($A_\text{X}$; see Section S4 in SI). Elliott's formula yields an excellent fit to the experimental data, with an absorption edge centered at the electronc bandgap $E_\text{g} = 2.360\pm0.001$\,eV (Fig. \ref{fig:Fig1}c). The exciton binding energy extracted from the fit, $E_\text{B}=31.6\pm0.9$\,meV, is higher than previous experimental estimates \cite{Galkowski2016} but lower than theoretical predictions \cite{Bokdam2016}. Based on the fit, the exciton energy is calculated as $E_\text{X}=E_\text{g}-E_\text{B}=2.328\pm0.002$\,eV ($532.6$ nm).
\medskip

At low fluences of the optical excitation, the exciton emission is responsible for the broad peak in the photoluminescence (PL) spectrum ($\text{FWHM}=109$\,meV) from the bare FAPB shown in Fig. \ref{fig:Fig1}d. The Stokes red-shift of the exciton emission relative to the absorption peak in Fig. \ref{fig:Fig1}c is in good agreement with previous reports \cite{Hu2021,Geng2023}). This emission is only weakly affected by exposure to temperatures of 150\,$^\circ$C for a period of 60 minutes after the PMMA coating, confirming that the sample is well protected against further fabrication steps, such as plasma-enhanced chemical vapor deposition (PECVD) of the top DBR.

\begin{figure*}
  \includegraphics{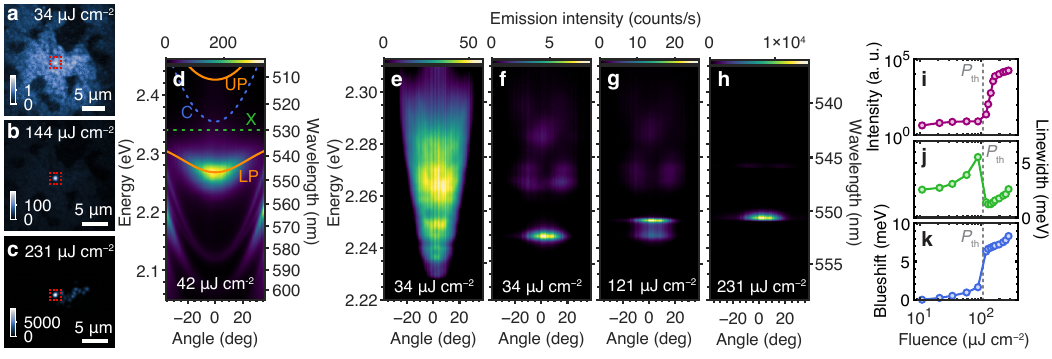}
\caption{\textbf{a--c,} Real-space emission from the microcavity, for excitation fluences below (a) and above (b, c) the polariton condensation threshold. \textbf{d--e,} Angle-resolved PL spectra for large excitation area. In (d) solid orange curves mark lower (LP) and upper (UP) exciton-polariton energy dispersions fitted to the spectra. Dashed green and blue curves: initial, non-hybridised exciton and cavity mode branches. \textbf{f--h,} Angle-resolved PL spectra for increasing excitation fluence from a smaller region marked in (a--c) with red dotted line. \textbf{i--k,} Maximum emission intensity, spectral linewidth and spectral peak position as a function of excitation fluence.}
  \label{fig:Fig3}
\end{figure*}

\subsection{Random photon lasing} 
The characteristics of the PL of bare FAPB change drastically for excitation fluences exceeding ${\sim}408$\,$\upmu$J/cm$^2$ (Fig.\,\ref{fig:Fig2}). The FWHM of the emission peak decreases from $122$ to ${\sim}18$\,meV, while the maximum emission intensity increases by at least an order of magnitude, as seen in Fig. \ref{fig:Fig1}e. This transition is attributed to the onset of random photonic lasing in bare FAPB, which arises due to the high optical gain of FAPB and strong multiple scattering at the boundaries between the crystalline grains of the random gain medium \cite{Sapienza_NatRevPhys2019}, as schematically depicted in Fig. \ref{fig:Fig1}a. Random lasing has been previously observed in other solution-based polycrystalline perovskite thin films, which points to the potential application of such easy-to-fabricate materials as low-cost semiconductor lasers \cite{Kao_Nanotechnology2021,Wang_Micromachines2022}.
\medskip

The onset and evolution of the random laser emission with increasing fluence of the optical pump illuminating a large (${\sim}40~\upmu\mathrm{m}\times40$ $\upmu$m) area of the sample is further characterized in Fig. \ref{fig:Fig2}. The real-space images of the PL below and above the lasing transition (see Fig. \ref{fig:Fig2}a--c) show that the spatial mode of the random laser is delocalized and speckle-like \cite{Wiersma2004}, in contrast to Anderson-localized modes in extremely strongly scattering materials \cite{Jiang2000}.
\medskip

Above the random lasing threshold, the broad-angle, narrow linewidth persists in the whole range of tested pump fluences, and the emission intensity continues to grow, as seen in Fig. \ref{fig:Fig2}d--j. However, the emission peak red-shifts with increasing fluence, see Fig. \ref{fig:Fig2}k, with a maximum red-shift of $12$\,meV reached at the maximum fluence of $1225$ $\upmu$J/cm$^2$. This effect can be attributed to the increase of the density of free carriers in the system, and thus the increase in overall Coulomb screening within the system. The system transitions to a degenerate electron-hole plasma and the refractive index of the medium grows \cite{Schlaus2019,Roeder2018}, leading to saturation of the gain and a red-shift in emission profile. This observed red-shift is consistent with a Drude-like response in semiconductor lasers well above the lasing threshold, such as ZnO \cite{Wille2016} and CsPbBr$_3$ \cite{Schlaus2019}.

\subsection{Strong light-matter coupling regime}

The strong light-matter coupling regime is achieved when the PMMA-encapsulated FAPB perovskite film is embedded in the full DBR microcavity  \cite{Su2017,Su2018,Su2021,Bouteyre2019,Liu:23,Laitz2023}, as shown in Fig. \ref{fig:Fig1}b. This regime is characterized by a spatially delocalized and strongly angle-dependent PL for large-area (${\sim}40\times40$ $\upmu\mathrm{m}^2$) illumination by a low-fluence pump, as shown in Fig. \ref{fig:Fig3}a, d, e. The angle-dependent emission can be fitted to a typical parabolic exciton-polariton dispersion shown in Fig. \ref{fig:Fig3}d. The emission is localized to the low-energy states of the lower polariton (LP) branch resulting from the avoided crossing between the bare exciton resonance ($E_\text{X}$) and the blue-detuned cavity photon with the energy $E_\text{C}>E_\text{X}$ (see Fig. S4 in SI), with the effective cavity photon mass $m_\text{C}=1.4\times 10^{-5}\,m_\text{e}$, where $m_\text{e}$ is the free electron mass. The LP mass in all samples is determined to be $m_\text{LP}\sim 10^{-4}\,m_\text{e}$, with the paticular dispersion highlighted in Fig. \ref{fig:Fig3}d corresponding to $m_\text{LP}=8.5\times 10^{-5}\,m_\text{e}$. Correspondingly, the thermal de Broglie wavelength of polaritons within this range of the effective mass is $\lambda_\text{dB}\sim 1$\,$\upmu$m, which is commensurate with the lateral size of the crystalline grains. Similarly to other perovskites \cite{Wang2018,Su2017,Fieramosca2019,Su2018,Polimeno2020}, FAPB exhibits large Rabi splitting in the range between $122$ and $256$ meV, which quantifies the strength of light-matter coupling in this material.
\medskip

A close-up of the lower polariton emission presented in Fig. \ref{fig:Fig3}e shows that a considerable population of polaritons in near-zero in-plane momentum states, corresponding to small angles, is distributed across a substantial range of energies. Spatial filtering of the low-fluence emission from a small area of the sample (${\sim}2\times2$\,$\upmu\mathrm{m}^2$) marked in Fig. \ref{fig:Fig3}a reveals that a significant proportion of the emission seen in Fig.\,\ref{fig:Fig3}e comes from the polaritons that are strongly confined \cite{Baumberg2018} in an effective `zero-dimensional' trap -- a `polaritonic dot' \cite{EPLandscape}. These traps are formed by the local variation of the microcavity length due to the surface roughness of the FAPB film encapsulated by PMMA layer. Although the RMS roughness of the encapsulated perovskite is small ($8-9$\,nm), the variation measured by contact profilometry can reach $60-100$ nm at the $0.16$\,$\upmu$m lateral scale (Fig. S2 in SI). This translates to a variation of the refractive index of the microcavity on the same lateral scale, and therefore to the formation of an effective photonic trap for the polaritons \cite{EPLandscape} (see Fig. S5 in SI). The majority of the polariton emission originates from the ground state in such an effective trap, as seen in Fig. \ref{fig:Fig3}f; however, emission from higher-order trapped states is non-negligible. Wide variation of the sizes and depths of the effective traps across the sample results in variation of the ground and excited state energies, which contribute to the broadband emission shown in Fig. \ref{fig:Fig3}e when a large area of the sample (${\sim}40\times40~\upmu\mathrm{m}^2$) is illuminated and the emission is spatially unfiltered.

\subsection{Polariton lasing}

Transition to polariton condensation and lasing in the microcavity sample occurs with increasing pump fluence, above ${\sim}121$ $\upmu$J/cm$^{2}$. The transition is signalled by spatial localization of the PL to the bright, randomly distributed sub-$\upmu$m polaritonic dots on the sample, as seen in Figs. \ref{fig:Fig3}b, c and Fig. S6 (SI). Spatially filtered, angle-resolved emission (Figs. \ref{fig:Fig3}g, h) from a single isolated dot shows that the condensate forms in the ground state of the effective polariton trap (SI Fig. S5). The polariton condensation is further characterized by highly nonlinear growth of the emission intensity above the threshold (Fig. \ref{fig:Fig3}i), linewidth narrowing (Fig. \ref{fig:Fig3}j), and the transition from unpolarized emission to lasing with a high degree of linear polarization (Fig. S7 in SI). Notably, the threshold for the onset of polariton lasing in FAPB is approximately four times lower than that for the random photon lasing in this material (cf. Figs. \ref{fig:Fig2}i and \ref{fig:Fig3}i).
\medskip

\begin{figure}
\begin{center}
  \includegraphics{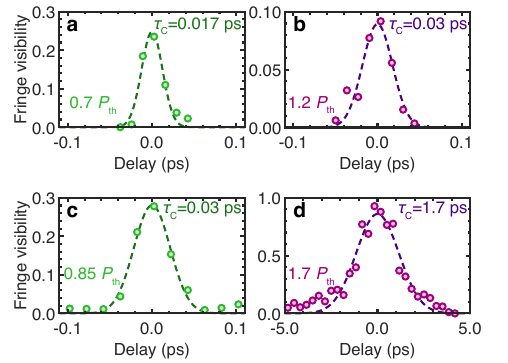}
 \end{center}
\caption{Fringe visibility as a function of time delay between the arms of the interferometer: \textbf{a, b} for bare perovskite film below (a) and above (b) the lasing threshold; \textbf{c, d,} for the microcavity sample below (c) and above (d) the polariton condensation threshold. Gaussian fits corresponding to $1/\text{e}$ coherence times $\tau_\text{c}$ are marked with dashed lines.}
  \label{fig:FigCo}
\end{figure}

In contrast to the random photon laser, the peak emission energy of the polariton laser experiences a blue-shift with increasing pump fluence, as seen in Fig. \ref{fig:Fig3}k. In the presence of polariton confinement, this behavior can be attributed to interaction of polaritons with the excitonic reservoir injected by the pump, as well as to polariton-polariton interactions \cite{Ferrier2011,Brichkin2011}. Both of these effects can be responsible for the broadening of the polariton laser linewidth observed for increasing pump fluences in Fig. \ref{fig:Fig3}j; careful delineation between the two mechanisms, which would allow us to draw conclusions regarding the polariton-polariton interaction strength in this material, requires further studies. Although broadened at large pump fluences, the polariton laser emission displays an order of magnitude smaller linewidth compared to the random photonic laser. 
\medskip

The range of spatial coherence of the polariton laser is limited by the size of the condensate, i.e., by the size of the polaritonic dot. However, a stark difference between the polaritonic and photonic lasing regimes is demonstrated by comparing the first-order temporal coherence, $g^{(1)}(\tau)$, of the corresponding lasers (see Section S6 in SI for more details). As seen in Fig. \ref{fig:FigCo}a, b, the coherence time of the random photonic laser, $\tau_c = 0.03$\,ps, extracted from the fringe visibility for varying time delay between the Michelson interferometer arms, is only marginally longer than the coherence time of a bare FAPB exciton ($0.017$\,ps) and is the same as the coherence time of polaritons below the condensation threshold, Fig. \ref{fig:FigCo}c. Remarkably, the polariton laser operating at a modest $1.7$ times the threshold pump fluence exhibits an improvement in the coherence time of two orders of magnitude to $\tau_c = 1.7$\,ps, as seen in Fig. \ref{fig:FigCo}d.
\medskip

In addition to the first-order temporal coherence, the second-order correlation function of the polariton laser emission at zero time delay, $g^{(2)}(\tau=0)$, displays a ${\sim}10 \%$ bunching effect close to the condensation (lasing) threshold,  approaching the value for a coherent laser $g^{(2)}(\tau=0)=1$ when $P>P_\text{th}$ (Fig. S10 in SI). This is in agreement with previous observations for a different material system \cite{Dang2008}. At the same time, we do not detect bunching below threshold, for $P<P_\text{th}$, 
as the coherence time $\tau_c$ of `thermal' polaritons is significantly shorter than the time resolution of the measurement setup (see Section S6 in SI).

\begin{figure*}
  \includegraphics{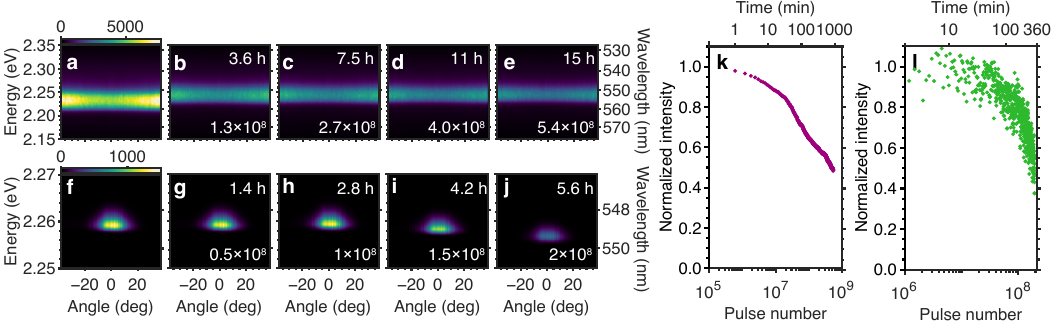}
  \caption{\textbf{a--e,} Angle-resolved random lasing emission spectra for bare FAPB film, for increasing number of excitation laser pulses (as indicated), at a fluence $\sim$10 times that of the lasing threshold. Measured for a large (${\sim}40~\upmu\mathrm{m}\times40$ $\upmu$m) area of the sample. \textbf{f--j,} Same, for polariton lasing of FAPB film in microcavity, at a fluence $\sim$2 times that of the lasing threshold. \mbox{\textbf{k, l,}}\,Random and polairton lasing intensity as a function of number of excitation laser pulses, for bare FAPB and microcavity-embedded FAPB films, respectively.}
  \label{fig:Fig4}
\end{figure*}

\subsection{Photostability of FAPB lasers}

The stability of perovskites against exposure to ambient conditions and light is a critical factor for potential applications \cite{Sanvitto2016,Su2021}. Both encapsulated perovskite samples produced and investigated here -- the bare samples protected by PMMA, and the perovskites embedded in microcavities -- showed remarkable durability. We observed little-to-no degradation of the encapsulated perovskite films, even after weeks of exposure to air. 
\medskip

The transition to random photon lasing in bare FAPB samples and to polariton lasing in the microcavity samples is driven by a pulsed optical pump, which means that the stability of the samples against continuing exposure to such laser pulses above the condensation and lasing threshold needs to be established. Fig.\,\ref{fig:Fig4}a--e, k show that the intensity of the random laser emission for bare FAPB decreases by approximately a factor of 2 after continuous exposure to the pump pulses at 10\,$P_{\text{th}}$ for ${\sim}15$ hrs (at 10\,kHz repetition rate). This is accompanied by a slight blue-shift, similar to previous reports \cite{Jia2017}.
\medskip

In contrast, the intensity of the polariton laser emission for FAPB embedded in a microcavity decreases by the same amount after continuous exposure to the pump pulses (at $\sim$2\,$P_{\text{th}})$ for a shorter period of ${\sim} 6$\,hrs (see Fig.\,\ref{fig:Fig4}f--j, l). However, the intensity remains relatively unaffected for approximately half of this time, see Fig. \ref{fig:Fig4}f--h. The decline in the laser intensity after this time is accompanied by a red-shift of the emission (Fig. \ref{fig:Fig4}i,j), with no direct correlation between emission intensity and energy.
\medskip

\section{Conclusion} 
To summarize, we produced a high-quality polycrystalline FAPB film with a constant thickness and minimal surface roughness. We embedded this solution-processed active material inside an optical microcavity using highly scalable methods of spin-coating (for both the FAPB film and the PMMA cavity spacers) and PECVD (for the DBRs). Both the FAPB film encapsulated by a protective layer of PMMA (`bare' FAPB) and full microcavity samples demonstrate remarkable stability in ambient conditions and do not exhibit significant degradation of optical properties when exposed to pulsed laser excitation for several hours.
\medskip

The FAPB embedded in a microcavity shows clear signatures of strong exciton-photon coupling and formation of exciton-polaritons. We demonstrated the transition to polariton condensation and lasing in the microcavity samples and characterized the coherence properties of the laser emission.
\medskip

Our polycrystalline FAPB films exhibit granulation and surface roughness on the sub-micrometer scale, which results in natural zero-dimensional traps for exciton-polaritons. The quantum confinement of exciton-polaritons inside these polaritonic quantum dots promotes bosonic condensation in the ground state of the traps. Compared to the (random) photon lasing that we also observed in bare FAPB at room temperature, the resulting polariton lasing is characterized by a lower threshold, an order of magnitude narrower linewidth, and two orders of magnitude longer coherence time. 
\medskip

The photostability of FAPB and the scalibility and efficiency of its solution-based integration into functional heterostructures demonstrated here makes this material promising for both fundamental and applied room-temperature polaritonics. Although the heterogeneous, fragmented nature of the condensates formed in the polycrystalline FAPB prohibits studies that rely on large areas of a uniform active material, such as monocrystalline CsPbBr$_3$ \cite{Su2017,Bao2022,BaoNMat2022}, the FAPB could be uniquely suited to applications that rely on zero-dimensional quantum confinement \cite{Liew2023}, such as polaritonic quantum light sources \cite{Volz2019,Imamoglu2019}.

\medskip
\section*{Supporting Information} \par 
Supporting Information is available from the Wiley Online Library or from the corresponding author.

\medskip
\section*{Acknowledgements} \par 
This work is funded by the Australian Research Council Centre of Excellence in Future  Low-Energy Electronics Technologies (CE170100039) and the Discovery Early Career Researcher Award (DE220100712). M.W. acknowledges support by Schmidt Science Fellows, in partnership with Rhodes Trust.

\medskip
\section*{Conﬂict of Interest} \par The authors declare no conﬂict of interest.

\medskip
\section*{Data Availability Statement} \par The data that support the ﬁndings of this study are available from the corresponding author upon reasonable request.

\bibliography{bib}

 \end{document}


\title{Robust Room-Temperature Polariton Condensation and Lasing in Scalable FAPbBr\textsubscript{3} Perovskite Microcavities:  Supporting Information}

\author{Mateusz Kr\'{o}l}
\thanks{These authors contributed equally}
\affiliation{ARC Centre of Excellence in Future Low-Energy Electronics Technologies and Department of Quantum Science and Technology, Research School of Physics, The Australian National University, Canberra, ACT 2601, Australia}

\author{Mitko Oldfield}
\thanks{These authors contributed equally}
\affiliation{ARC Centre of Excellence in Future Low-Energy Electronics Technologies and School of Physics and Astronomy, Monash University, Clayton, VIC 3800, Australia}

\author{Matthias Wurdack}
\thanks{These authors contributed equally}
\affiliation{ARC Centre of Excellence in Future Low-Energy Electronics Technologies and Department of Quantum Science and Technology, Research School of Physics, The Australian National University, Canberra, ACT 2601, Australia}
\affiliation{Hansen Experimental Physics Laboratory, Stanford University, Stanford CA 94305}

\author{Eliezer Estrecho}
\affiliation{ARC Centre of Excellence in Future Low-Energy Electronics Technologies and Department of Quantum Science and Technology, Research School of Physics, The Australian National University, Canberra, ACT 2601, Australia}
\author{Gary Beane}
\affiliation{ARC Centre of Excellence in Future Low-Energy Electronics Technologies and School of Physics and Astronomy, Monash University, Clayton, VIC 3800, Australia}
\author{Yihui Hou}
\affiliation{School of Engineering, The Australian National University, Canberra, ACT 2601, Australia}
\author{Andrew G. Truscott}
\affiliation{Department of Quantum Science and Technology, Research School of Physics, The Australian National University, Canberra, ACT 2601, Australia}
\author{Agustin Schiffrin}
\affiliation{ARC Centre of Excellence in Future Low-Energy Electronics Technologies and School of Physics and Astronomy, Monash University, Clayton, VIC 3800, Australia}
\author{Elena A. Ostrovskaya}
\affiliation{ARC Centre of Excellence in Future Low-Energy Electronics Technologies and Department of Quantum Science and Technology, Research School of Physics, The Australian National University, Canberra, ACT 2601, Australia}

\begin{abstract}
\end{abstract}

\maketitle

\vspace*{-3\baselineskip}
\section{Sample Fabrication}

Chlorobenzene (CBZ, anhydrous, 99.8\%), N,N-dimethylformamide (DMF, anhydrous 99.8\%,), dimethyl sulfoxide (DMSO, $\ge$99.9\%),  formamidinium bromide (FABr, >99.99\%), lead (II) bromide (PbBr\textsubscript{2}, >99.99\%), poly(methyl methacrylate) (PMMA, Mw $\simeq$ 996,000), and anisole (methoxybenzene, anhydrous 99.7\%) were purchased from Sigma-Aldrich and used as received.

\subsection{FAPbBr\textsubscript{3} synthesis} A precursor solution was formed by mixing FABr and PbBr\textsubscript{2} in their powdered form at a 1.03:1 ratio by molar weight, with a 3\% excess of FABr used to promote high crystal quality and minimize unreacted PbBr\textsubscript{2} \cite{Si2020}. Dimethyl sulfoxide (DMSO) and dimethylformamide (DMF) were added to the powder to form a 0.7 mol solution that was shaken for 60 minutes until the solids were completely dissolved. Next, 30 $\upmu$L of the solution was deposited on the bottom DBR stack on a fused silica (SiO\textsubscript{2}, FS) substrate that was inside a spin-coater. The DBR/FS substrates were washed in acetone, isopropanol and ethanol for 20 minutes, dried with N\textsubscript{2} and treated in a UV-ozone plasma cleaner for 10 minutes.
\medskip

Samples were spun at 4000 rpm for 40 s using a 1000 rpm/s ramp rate. Chlorobenzene was used as the antisolvent and 120 $\upmu$L of CBZ was dropped onto the solution after 7.5 seconds to aid crystallization. The samples were then annealed on a hot plate at 80\,$^\circ$C for 20 minutes. 

\medskip
During spin-coating, the rapid rotation of the sample leads to evaporation of the solvents. This increases the concentration of the remaining solution and leads to the formation of seed crystals, where multiple monomers -- in our case, suspended lead salt-formamidinium salt complexes -- may nucleate to form single crystals \cite{DeYoreo2015}. In our process, heterogeneous nucleation occurs, where the monomers begin to crystallize at an interface, such as on the substrate or on particles suspended in solution (e.g. undissolved salts or contaminants). 

\subsection{PMMA deposition} 
A 4\% molar weight solution of poly(methyl methacrylate) (PMMA) in anisol was deposited onto the FAPB/DBR sample and spun at 3000 (6000) rpm for 40 (60) seconds with a ramp rate of 500 rpm/s. The samples were then annealed on a hot plate at 90\,$^\circ$C for 120 seconds. We note that, while FAPB is stable up to the temperatures of 300\,$^\circ$C, PMMA has a melting point of 160\,$^\circ$C. This has to be taken into account both in the process of thermal annealing and in subsequent exposure of the sample to the pump laser, which produces localized heating.

\subsection{DBR fabrication} The top and bottom DBR stacks were fabricated using plasma-enhanced chemical vapour deposition at 150\,$^\circ$C, with an RF driving frequency of 50 MHz. The DBR stacks were fabricated with alternating $\lambda/4$ layers of silicon nitride (SiNx) and silicon oxide (SiOx), to satisfy the Bragg condition for the central wavelength of the Fabry-P\'{e}rot resonator. The chamber pressure was set to 650\,mTorr (86.7\,Pa, 8.55 $\times$ 10\textsuperscript{-4} atm).

\medskip
For SiNx deposition, 980 sccm of N\textsubscript{2} gas, 14 sccm of NH\textsubscript{3} gas and 18 sccm of SiH\textsubscript{4} gas were introduced at a low frequency power of 20 W. For SiOx deposition, 161 sccm of N\textsubscript{2} gas, 710 sccm of N\textsubscript{2}O gas and 9\,sccm of SiH\textsubscript{4} gas were introduced at the same low frequency power of 20 W.

\medskip
Prior to depositing the full DBR stacks onto the substrates, growth rates and the refractive indices of the SiNx and SiOx layers were determined separately via two stages of ellipsometry measurements. The first stage measured the thickness and refractive index of SiNx and SiOx after 5 minutes of deposition. The time required for a layer of $\lambda/4$ thickness was calculated from both the deposition time and the measured refractive index. The second stage then measured the achieved thickness and refractive index versus the desired thickness. The times for deposition were then calibrated accordingly.

\medskip
The bottom DBR stack was fabricated on fused silica with 14.5 SiNx/SiOx mirror pairs. After deposition of FAPB and PMMA via spin-coating, as described above, the top DBR stack needed to complete the cavity was fabricated with 11.5 SiNx/SiOx mirror pairs.

\section{Surface roughness characterization}

\begin{figure}[h]
\centering
  \includegraphics[width=0.35\textwidth]{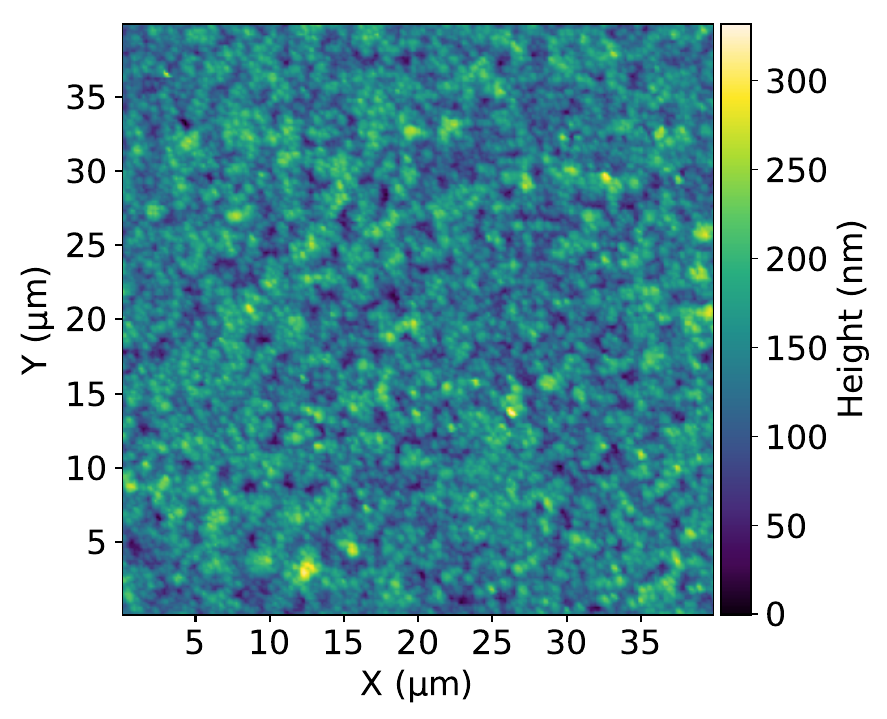}
\caption{Atomic force microscopy scan of a 500 nm thick FAPB thin film.}
\end{figure}

\begin{figure*}[h]
\centering
  \includegraphics[scale=1.25]{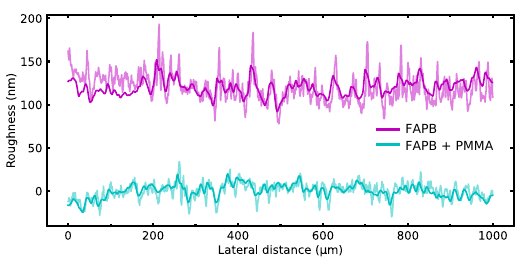}
\caption{Surface roughness for two different FAPB thin films, where the magenta curve is for FAPB only and the cyan curve is for FAPB covered with PMMA. The faint background traces show a scan over the same section with a higher scanning resolution (point spacing of 0.16\,$\upmu$m versus 1.11\,$\upmu$m). The traces have been offset by 80\,nm for clarity.}
  \label{fig:FigProfilometer}
\end{figure*}

\section{Experimental setup}

\begin{figure*}[h]
\centering
  \includegraphics[]{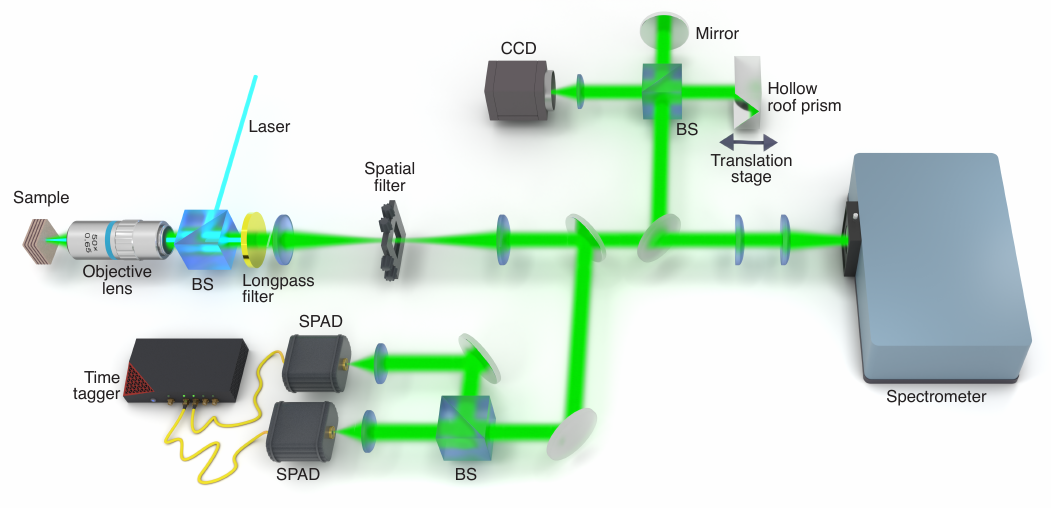}
\caption{Schematic of the experimental setup. BS -- beam splitter, SPAD -- single-photon avalanche photodiode.}
  \label{fig:ExpScheme}
\end{figure*}

Schematic of the experimental setup is presented in Fig.\,\ref{fig:ExpScheme}. For the optical pump, a laser with a wavelength of 1030\,nm (1.204\,eV) (Light Conversion Carbide, 6\,W) with 250\,fs pulse duration was used. The nominal 200\,kHz repetition rate of the laser (used in most of the experiments) can be reduced by a built-in pulse picker (10\,kHz for long-term stability measurements presented in Fig. 5 of the main text). The output of the laser was either frequency doubled to 515\,nm (2.408\,eV) with a BBO crystal or converted to 482\,nm (2.572\,eV) with an OPA (Light Conversion Orpheus).
\medskip

The excitation beam was directed through a 90:10 beamsplitter and focused on the sample with a microscope objective lens with 50$\times$ magnification and numerical aperture of 0.65. The emitted light is collected through the same objective lens in reflectance configuration. The reflected excitation laser light was blocked by a long-pass filter. A lens with 150\,mm focal length was used to create an intermediate image of the sample, which can be spatially filtered using a tunable slit. 
\medskip

Spectroscopic measurements were performed using an Andor Shamrock spectrometer equipped with an Andor iXon EMCCD camera. Angular dependence was obtained by imaging the Fourier plane of the objective lens onto the entrance slit of the spectrometer.

\section{Elliott fit of absorbance spectrum}

The absorbance spectrum shown in Fig. 1c of the main text exhibits an absorption onset at $\sim$2.3 eV, above which we see a monotonic increase in absorption with increasing photon energy. Previous literature has employed models based on Elliot's theory, which assumes that the valence band and conduction band possess a parabolic dispersion \cite{Yang2017_bottleneck,Fu2017,Yang2015}. However, as outlined by Saba and Setsu \textit{et al.} \cite{Saba2014,Sestu2015}, a non-parabolic term must be introduced to accurately capture the dynamics of the system at photon energies well above the electronic bandgap, $E_\mathrm{g}$. As such, we employ Elliott's formula with an added term, $\xi$, which describes the non-parabolic behavior above bandgap \cite{Singh2016,Chang2022}. Our total absorbance, $A_\mathrm{tot}(\hbar\omega)$, is a sum of the charge carrier contribution, $A_\mathrm{CC}(\hbar\omega)$ and the exciton contribution, $A_\mathrm{X}(\hbar\omega)$

\begin{equation}\label{eq: Elliott full}
    A_\mathrm{tot}(\hbar\omega) = A_\mathrm{CC}(\hbar\omega) + A_\mathrm{X}(\hbar\omega),
\end{equation}

where the charge carrier contribution is given by

\begin{equation}\label{eq: Elliott A_CC}
    A_\mathrm{CC} = A_1 \frac{2\pi\sqrt{E_{\mathrm{B}}}}{\hbar\omega}\frac{1}{\Gamma}\int_{E_\mathrm{g}}^\infty\sech\left(\frac{\hbar\omega - E_\mathrm{CC}}{\Gamma}\right)\frac{1+\xi(E_\mathrm{CC}-E_\mathrm{g})}{1-\exp\left(-2\pi\sqrt{\frac{E_{\mathrm{B}}}{E_\mathrm{CC}-E_\mathrm{g}}}\right)} \textrm{d}E_\mathrm{CC},
\end{equation}

and the exciton contribution is given by

\begin{equation}\label{eq: Elliott A_EC}
    A_\mathrm{X} = A_2\frac{2\pi {E_{\mathrm{B}}}^{3/2}}{\hbar\omega}\frac{1}{\Gamma}\sum_{j=1}^{j_\text{max}}\frac{1}{j^3}\sech\left(\frac{\hbar\omega - E_\mathrm{g} + \frac{E_{\mathrm{B}}}{j^2}}{\Gamma}\right),
\end{equation}

where $E_{\mathrm{B}}$ is the exciton binding energy, $E_\mathrm{g}$ is the electronic bandgap, $\hbar\omega$ is photon energy, $\Gamma$ is the exciton bandwidth (i.e., standard deviation of exciton peak, where $\text{FWHM} = \gamma_\text{X} = 2\sqrt{2\ln{2}}\,\Gamma \approx 2.355\,\Gamma$), $j$ is the excitonic state (i.e., 1s is $j = 1$). A maximum value of $j_\text{max} = 7$ was chosen as it sufficiently captures all of the excitonic contributions to the absorption for our measurement \cite{Chang2022}. The non-parabolic term is captured by $\xi(E_\mathrm{CC}-E_\mathrm{g})$, which may be given by

\begin{equation}\label{eq: Elliott non-parabolic}
    \xi(E_\mathrm{CC}-E_\mathrm{g}) = 10\,R (E_\mathrm{CC} - E_\mathrm{g}) + 126\,R^2(E_\mathrm{CC}-E_\mathrm{g})^2,
\end{equation}
where $R = \frac{m_\text{e}^2}{\hbar^4}c_{np}$, with $c_{np}$ being the non-parabolic factor, and $m_\text{e}$ being the electron mass. We take $R \approx 0.11$ eV\textsuperscript{$-1$} from the work of Setsu \textit{et al.} \cite{Sestu2015}, where they determine the non-parabolic contribution for MAPbBr\textsubscript{3}, a similar perovskite. The values for $E_\mathrm{g}$ and $E_{\mathrm{B}}$, as well as for other quantities in Equations\,\eqref{eq: Elliott A_CC} and \eqref{eq: Elliott A_EC}, retrieved from the fit of our experimental absorbance data, are shown in Table \ref{table: Elliott fit values}. The exciton energy, $E_{\mathrm{X}}$, is taken as $E_\mathrm{g} - E_{\mathrm{B}}$.

\begin{table*}[h]
\caption{Values extracted from the fit performed in Fig. 1c using Elliott's theory with an additional non-parabolic term.}
\label{table: Elliott fit values}
\centering
\begin{tabular}{@{}c
>{\columncolor[HTML]{FAF1FF}}c c
>{\columncolor[HTML]{FAF1FF}}c c
>{\columncolor[HTML]{FAF1FF}}c c@{}}
\toprule
$A_1$         & $A_2$         & $E_\mathrm{g}$ (eV)      & $E_{\mathrm{B}}$ (meV)  & $E_{\mathrm{X}}$ (eV)      & $\Gamma$ (meV) & $R$ (eV$^{-1}$)\\ \midrule
0.28$\pm$0.03 & 0.62$\pm$0.02 & 2.360$\pm$0.001 & 31.6$\pm$0.9 & 2.328$\pm$0.002& 28.0$\pm$0.5   & 0.028$\pm$0.001 \\ \bottomrule
\end{tabular}%
\end{table*}

\newpage
\section{Additional characterization of polariton lasing}

\begin{figure*}[h]
\centering
  \includegraphics[width=0.32\textwidth]{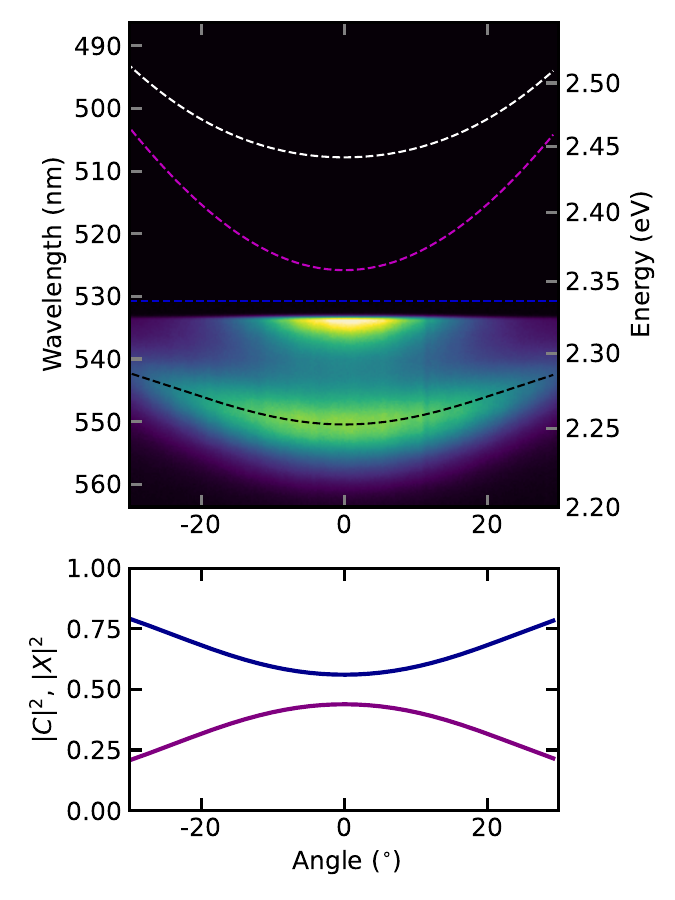}
  \includegraphics[width=0.32\textwidth]{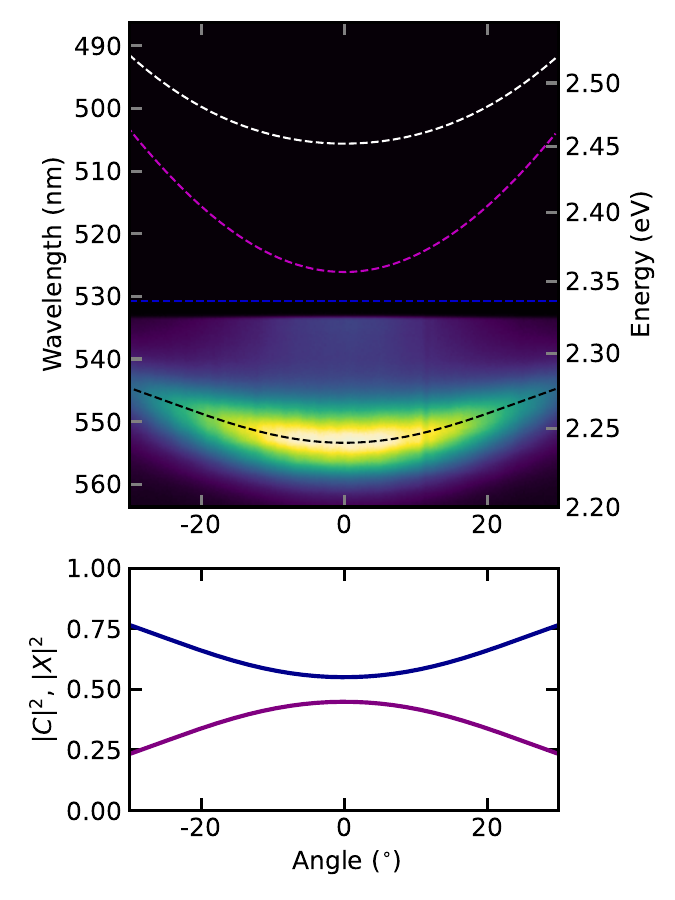}
\caption{Angle-resolved photoluminescence spectra at two different positions on the microcavity sample. The white (black) dashed traces correspond to the upper (lower) polariton branch. The dark blue trace corresponds to the exciton line, $E_X = 2.328$\,eV (532.6\,nm). The magenta trace corresponds to the cavity photon dispersion, $E_C$. Bottom panels show the corresponding excitonic ($\left|X\right|^2$, solid dark blue trace) and photonic ($\left|C\right|^2$, solid purple trace) Hopfield coefficients which determine the respective excitonic and photonic components of the polaritons. The edge of the emission around 532 nm is due to a long-pass filter.}
\end{figure*}

\begin{figure}[h]
\centering
  \includegraphics{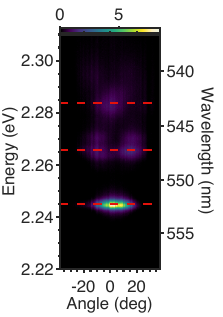}
\caption{Spatially-filtered angle-resolved photoluminescence spectrum from the microcavity sample. Red dashed lines mark three lowest energy levels for polaritons in an effective potential trap formed due to the surface roughness of the granular active material. The polaritons are formed by photon modes confined in a 0.97\,$\upmu$m wide and 98\,meV high photonic potential trap (effective mass of $1.65\times10^{-5} m_\text{e}$) strongly coupled with the excitonic resonance.}
\end{figure}

\begin{figure*}[h]
\centering
  \includegraphics[width=0.475\textwidth]{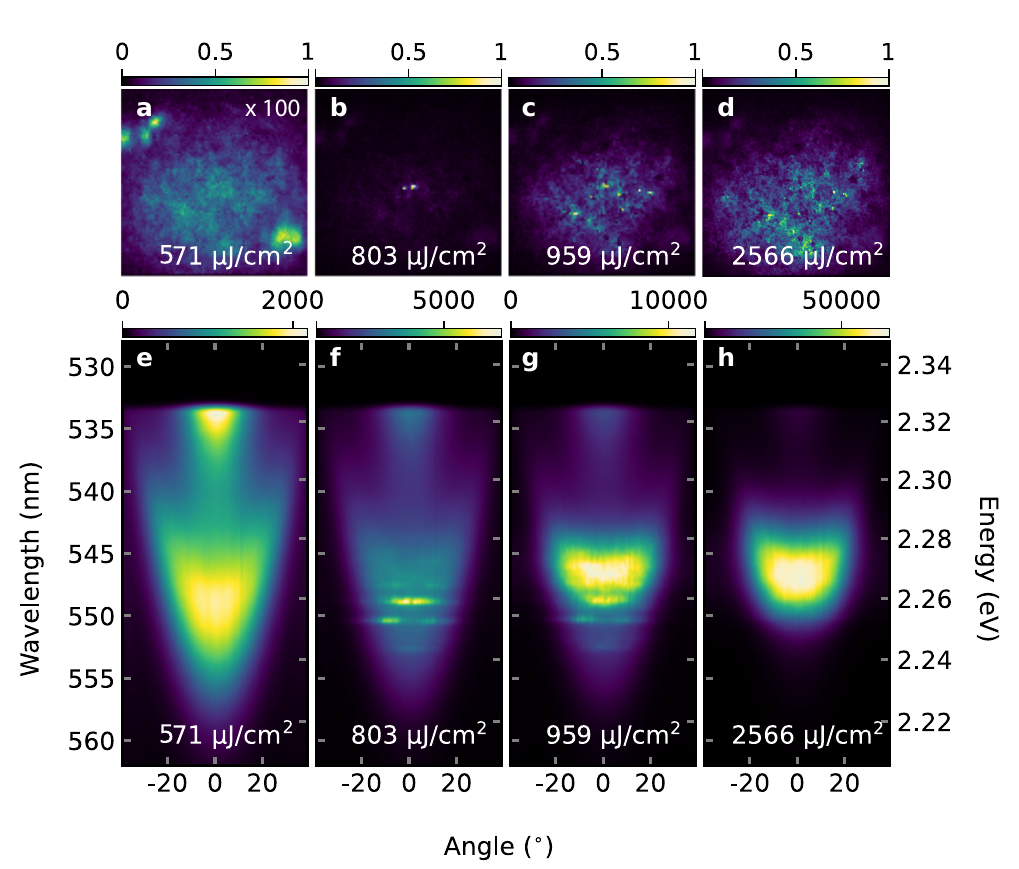}
  \includegraphics[width=0.475\textwidth]{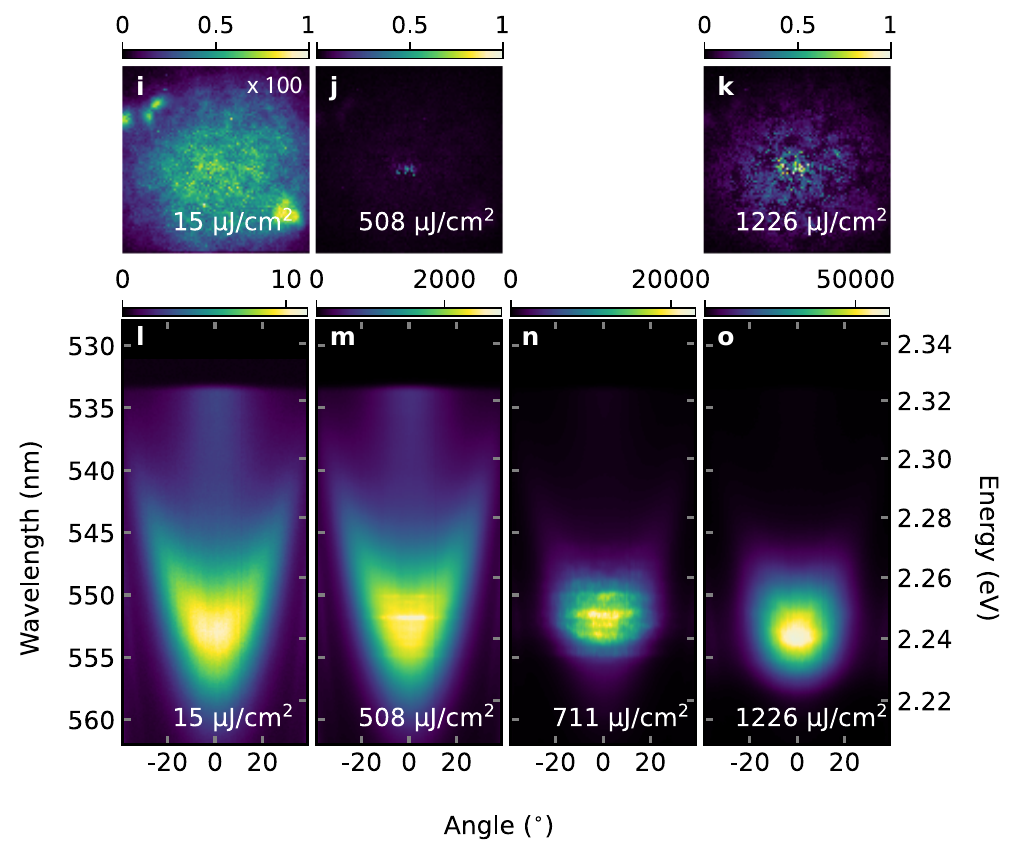}
\caption{Real-space (\textbf{a--d}, \textbf{i--k}) and angle-resolved (\textbf{e--h}, \textbf{l--o}) photoluminescence measurements showing the evolution of a polariton condensate as a function of incident pump fluence for large-area illumination ($\sim$80 $\upmu$m $\times$ 80 $\upmu$m sampling window) of the \ch{FAPB} microcavity. Panels \textbf{a--h} and \textbf{i--o} correspond to two different positions on the same sample. }
\end{figure*}

\begin{figure*}[ht]
\centering
  \includegraphics{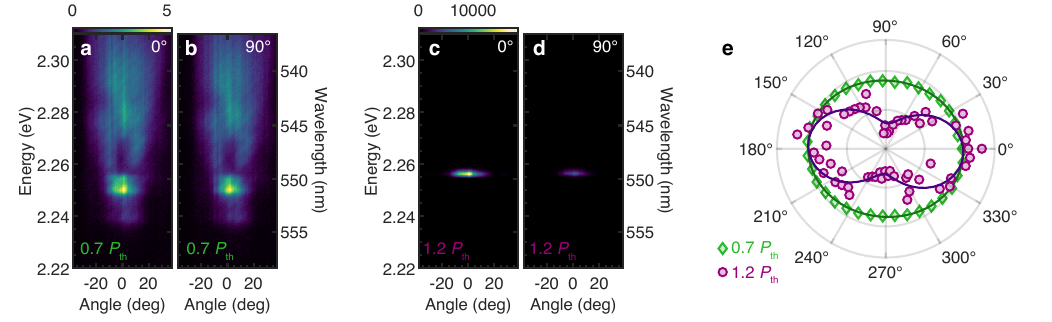}
\caption{\textbf{a--d,} Angle resolved photoluminescence spectra below and above the condensation threshold detected in two orthogonal linear polarizations. \textbf{e,} Normalized emission intensity dependence on the polarization angle below and above the condensation threshold. }
\end{figure*}

\newpage
\section{Temporal coherence measurements}
\begin{figure*}[h]
\centering
  \includegraphics{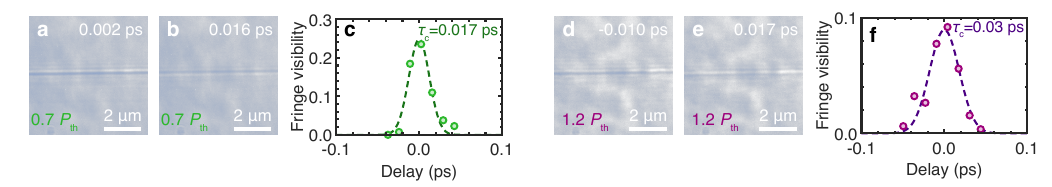}
\caption{Temporal coherence in bare FAPB film. \textbf{a, b,} Interferograms at two different time delays between the arms at excitation fluence below the lasing threshold. \textbf{c,} Fringe visibility versus delay between the arms. Experimental results are fitted with Gaussian function. \textbf{d, e,} Interferograms at two different delays between the arms at excitation fluence above the lasing threshold. \textbf{f,} Fringe visibility versus delay between the arms. Experimental results are fitted with Gaussian function.}
  \label{fig:FigCoherenceBare}
\end{figure*}

\begin{figure*}[h]
\centering
  \includegraphics{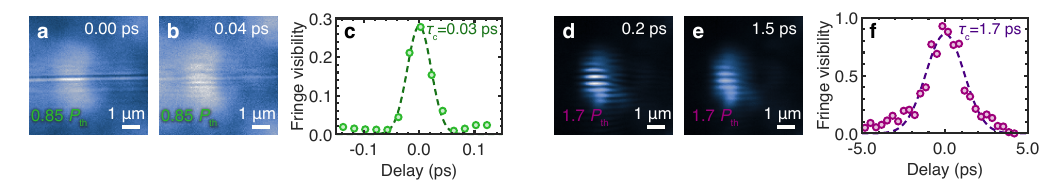}
\caption{Temporal coherence in microcavity sample. \textbf{a, b,} Interferograms at two different delays between the arms at excitation fluence below the condensation threshold. \textbf{c,} Fringe visibility versus delay between the arms. Experimental results are fitted with Gaussian function. \textbf{d, e,} Interferograms at two different delays between the arms at excitation fluence above the condensation threshold. \textbf{f,} Fringe visibility versus delay between the arms. Experimental results are fitted with Gaussian function.}
  \label{fig:FigCoherenceCavity}
\end{figure*}

The temporal coherence measurements (Figs.\,4, \ref{fig:FigCoherenceBare} and \ref{fig:FigCoherenceCavity}) were performed using a Michelson interferometer (see Section S3). One arm of the interferometer was composed of a stationary silver mirror. The second arm was composed of a hollow roof prism placed on a translation stage. The hollow roof prism inverts the image along the horizontal direction and the position of the translation stage allowed for precise control over time delay between the two interferometer arms. Finally, light from both interferometer arms was combined and imaged on a camera using a concave lens with the focal length of 1000\,mm. 
\medskip

\begin{figure}[h]
\centering
  \includegraphics{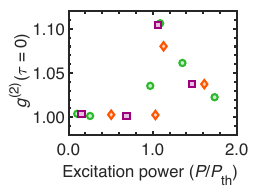}
\caption{Measured second-order correlation function at zero time delay as a function of excitation fluence relative to the condensation threshold. Different markers show results obtained at different positions on the sample.}
  \label{fig:g2}
\end{figure}

Photon correlation measurements used to characterize the second-order coherence, were performed using a Hanbury Brown--Twiss setup (see Section S3). The emitted light after the spatial filter was separated into two beams by a 50:50 beamsplitter. Both beams were then focused on two separate single-photon avalanche photodiodes (SPADs) (IDQ ID100). The timing of every single photon event detected with the SPADs was recorded with a time tagger (IDQ ID1000), and used to calculate the second-order correlation function presented in Fig.\,\ref{fig:g2}. The time resolution of the system of $\sigma = 51$\,ps originating from the jitter of the SPADs and precision of the time tagger was determined using the pump laser light reflected from the sample.

\clearpage
\newpage
\bibliography{bib}